# EFFICIENT ALGORITHMS TO ENHANCE RECOVERY SCHEMA IN LINK STATE PROTOCOLS

Radwan S. Abujassar[1] and Mohammed Ghanbari[2]

[1]School of Computer Science and Electronic Engineering, University of Essex, Colchester, UK
`rabuja@essex.ac.uk`
[2]Essex, Wivenhoe Park, CO43SQ
`ghan@essex.ac.uk`

## ABSTRACT

*With the increasing demands for real-time applications traffic in net- works such as video and voice a high convergence time for the existing routing protocols when failure occurred is required. These applications can be very sensitive to packet loss when link/node goes down. In this paper, we propose two algorithms schemas for the link state protocol to reroute the traffic in two states; first, pre-calculated an alternative and disjoint path with the primary one from the source to the destination by re-routing traffic through it, regardless of the locations of failure and the number of failed links. Second, rerouting the traffic via an alternative path from a node whose local link is down without the need to wait until the source node knows about the failure. This is achieved by creating a new backup routing table based on the original routing table which is computed by the dijkstra algorithm. The goal of these algorithms is to reduce loss of packets, end-to-end delay time, improve throughput and avoiding local loop when nodes re-converge the topology in case of failure.*

## 1. INTRODUCTION

Internet Protocol (IP) can use various types of routing protocols, such as, Link State (LS) and Distance Vector Routing Protocols (DSDV). The LS Protocol has been successful over the past few years, because it provides networks with numerous optimization techniques which lead to fast convergence enhancements. In case of failure, the LS protocol needs to re-update the routing table to divert the failure affected traffic along another path around the failure to destination which takes seconds. In real time applications such as VOIP, the fast rerouting mechanism is considered one of the most important solutions for alleviating congestion, as well as reducing delay and the packet loss in the network when routers start to re-converge the topology after the topology has been changed due the link/node failure. The recovery mechanism comes with appropriate solutions to avoid local loop in the network between nodes by finding an alternative backup path for delivering packets to their destination. By pre-computing an alternative backup path, traffic can pass through it in case any link on the primary path fails without having to wait for updating of the routing table. The main purpose of the IP fast re-route mechanism is to address the slow convergence time networks when failure occurs in the network. Currently, the convergence time is 100's of a millisecond or even 10's of seconds in the Border

Getaway Protocol (BGP) network, which is unsatisfactory [9]. Each node associated with a failure needs to re-compute a new shortest path, update the routing table via the routing protocol and then propagate these updates for all nodes associated with the malfunction. In addition, the recovery mechanism has to address the period spent upon the problem with respect to the time required to detect the failure and compute the new shortest path. The Open Shortest Path First (OSPF) routing protocol is used as a dynamic link state protocol for TCP/IP or UDP traffic and is designed to update topological information by sending a Link State Advertisement (LSA) based on the presence of a failure. Table 1 illustrates the default and minimum times for the routing protocol to re-compute a new shortest path.

| Timer | Default Value | Minimum Value |
|---|---|---|
| Notification timer | 2s | 10ms |
| Link state Packet (LSP) | 50ms | 1ms |
| Shortest path computation timer | 5.5s | 1ms |
| Processing phase Typical | Processing phase Typical values | |
| LSP processing | 10ms/hop | |
| SPF computation | 100 - 400 ms | |
| Forwarding information update | 20 entries / ms | |

Table 1: COMPONENTS OF THE FAILURE RESTORATION TIME [11]

Micro-loops can occur between nodes, however, which do not receive notification when a failure occurs. The effect of this can lead to increasing the utilization of the links, which may exceed the link capacity. In case of failure, the source will keep sending packets until it receives a notification that a failure has occurred. In this paper, we pro- pose and describe two new algorithms that can create a backup routing table with a loop back free in the network. The first algorithm is called an Alternative Routing Table Full Path (ARTFP), which re-routes the traffic from source to its destination through an alternative disjoint path to the primary one. This backup path contains a backup next-hops not connected with any node on the primary path. The second algorithm is called an Alternative Routing Table Connected Path (ARTCP), which re-routes the traffic from the node where it is connected directly to the failure on the primary path, and it can reroute the traffic through the backup routing table to the destination. We have concentrated on the original routing table, which is constructed by the LS protocol to create the backup routing table in order to find all possible alternative routes to the destination excluding all primary routes between source and destination for each node on the topology.

This paper is organized as follows: Section 2 discusses the related work for ART algorithms. Section 3 explains the mechanism of the algorithms. Section 4 introduces the basic concept for the ARTCP and ARTFP algorithms in details. Section 5 explains the result and section 6 concludes how ART algorithms can be improved and the direction of any future work.

## 2. RELATED WORK

Traditional routing protocol schemes allow traffic to pass via a single shortest path. In current networks, nodes or link failures produce factors that cause a disruption to the flow of traffic until the routing protocol recalculates their routing tables and computes a new primary path to the destination. An alternative path aims to alleviate these disruptions by making the source node pass the traffic through it when the primary path goes down. There are two types of recovery mechanisms: the protection and restoration. The protection schema is a proactive mechanism, which calculates backup routes in advance while the restoration schema is reactive by calculating the backup routes when failure has been detected. The restoration schema considers more flexibility with regard to the location of failures, because the recovery mechanism will take action based on their locations. The disjoint path between source and destination considers a best solution to recover the network regardless of the location of the failure and the number of nodes affected by the malfunction. This kind of mechanism guarantees that the traffic arrives through it to the destination with loop free in the network [8]. In [16] the author discusses the basic K-Shortest Path First (K-SPF) algorithm and hoe to compute multiple paths to the destination. In [22] the algorithm design has shown and demonstrated how to compute the set of K-shortest paths. Hence, the algorithm has shown that it does not consider hop-by-hop forwarding to establish the ingress and egress router. Several algorithms have been proposed to solve the K-shortest path problem. In addition, most solutions do not give credence to the hop-by-hop multipath and it is considered not applicable for the first disjoint hop. In [14], a new routing technique was introduced to alleviate packet loss when links fail. This technique is based on alternative next hop counters that allow the routers to find alternative paths and re-direct the traffic in case of failure. The main concept involved with this technique is to add a counter to the packet header after the shortest path has been computed by the protocol and then pass the packets along the right path. The Open Shortest Path First protocol (OSPF) based on the Dijkstra algorithm computes the shortest path. The minimum path cost will be determined by comparing it with the other candidate paths [15]. The routing protocol will re-route packets from a backup path when any link on the primary one fails. There are two kinds of the Dijkstra algorithms. Firstly, there is a Dijkstra algorithm to compute the best path by removing the links with bandwidths less than a threshold. Secondly, there is an on demand Dijkstra algorithm, which generates the shortest path tree to the pre-computation node [3, 5] .The node, will be added to the tree depending on the bandwidth request. However, IP recovery emphasizes two cases. First, the time required to detect failure. Second the time to compute the shortest path. Recently, many techniques have been published that contribute to improve recovery time. In [19] the author mentions how recovery of the network can be achieved from failure within a short time. The aim of IP recovery is to offer a protection mechanism with loop free in the network. The loop in the network forms one of the main problems with some of the existing techniques. In [12], a new fast IP re-route mechanism for Shared Risk Link Group (SRLG) has been proposed without the need for assumptions. The author bases this mechanism on message advertisement between nodes to identify which node on the topology is able to re-route the traffic around the failed link. In [21] the author proposed a new mechanism, termed Failure Insensitive Technique (FIR). FIR uses the specific forwarding interface to provide a backup next hop without local loop. The FIR mechanism makes the node that is connected to the failure adds a new header by re-encapsulating the packets and then re-sending them to adjacent nodes to inform them of the failure through the interface packets that arrive. Hence, based on the interface packets when failure occurs, the adjacent nodes will re-route the affected packets. The other nodes will not know about the failure as packets are sent according to pre-computed routing tables. FIR has several drawbacks such as the encapsulation of packets is not desirable because that will reduce the throughput and make the end-to-end delay longer. In addition, FIR cannot provide protection against node failure. In [13, 7, 20], the Inter- net Protocol Fast Re-Route (IPFRR) is an applicable technique. It includes the Loop Free Alternate (LFA), U-turn and not-via address [10, 26].The drawback to the IPFRR technique is that loop free is not guaranteed because the packet can be returned to the source with regard to a specific forwarding pre-computed routing table for each node on the network. IPFRR mechanism attempts to keep the performance of the network under maintenance during the updating time when failure occurs. The important issue concerning the IPFRR technique is to maintain failure in the order of millisecond. In [4] the authors show IPFRR not-via address still lags by using a computational complexity. Therefore, the authors discuss a new idea that can improve this problem by presenting a lightweight not-via schema by making node to understand the redundant trees for IPFRR. The authors in [2, 18] created a new algorithm for generating alternative paths by constructing trees for each node, which are connected to its root. In the routing protocol, many extensions have been proposed to the link state and Intermediate System-Intermediate System (IS-IS) to improve restoration times, such as, activated link state messages to report failures in the network, compute the second shortest path and traffic engineering to balance the load through many paths for the same cost[24,17]. In [25] the authors propose a new schema to provide protection for most of links. Less Tunnel Server (LTS) algorithms use a few backup paths for link protection after they explore the shortest path. LTS algorithms use only a small number of shortest path calculations and exclude all links under the protection of IPFRR mechanisms.

# 3 PROPOSITION

## 3.1 PROBLEM ANALYSIS

Node/Link failure in the network will lead to produce a sequence of disruptions to de- liver the traffic to the destination until the routing protocol re-converge the new topology with the new routing table. However, the packets reaching the failed component could be suffering loops or they will be dropped. When failure occurs, the routers will take a few milliseconds until they can detect it at the physical layer up to several 10's of sec- onds. This amount of time will lead to packets becoming unavoidably lost. In addition, the time taken for each router on the topology to react when it knows about a failure requires it to generate and flood the new routing table with a new primary path to the destination. Hence, there are two new algorithms providing a mechanism for routers on the topology, which can rapidly invoke rerouting the traffic to the destination through an alternative path that is not affected by the failure.

## 3.2 ART MECHANISM

The originality of Alternative Routing Table (ART) mechanism lies in the method, which is used to find the backup path from the original routing table. This approach does not require any extensions on the router but it depends on the number of adjacent node for each node on the topology. We assume that each node has at least one adjacent node that is capable of acting as a backup in case of a failure. This will give each node a high possibility for it to anticipate the second shortest path regarding the original rout- ing table. The ART mechanism considers two cases, namely, re-routing from the source node which is called ART with Full Path (ARTFP) and re-routing from the node that is directly connected to the failure which called ART with Connected Path (ARTCP). The ART algorithms are proactive because they compute a new routing table including a backup path for all nodes on the topology in advance. The ART algorithms work until the backup routing table is complete. By using these algorithms, all possible paths to the destination are computed. The algorithm is executed into five steps:

### Algorithm Steps

1. All nodes send packets to all adjacent nodes not connected to the primary path on the topology. The nodes will receive packets to check if there are any possible alternative routes to the destination not connected to the primary path.

2. If the answer coming from an adjacent node is "No", then the algorithm will send another packet to enquire if there is any route from the adjacent node to the adjacent source node.

3. When the source receives a "Yes" answer from the adjacent node, it should select a loop free node next hop then will add this node as first backup next hop in case of failure.

4.  If the node adjacent to the source has a disjoint route to the destination then the node will add her adjacent node as a first next backup hop and her neighbor as a second backup hop.

5.  ART algorithms will repeat these steps until the new backup routing table is completed for each node on the topology.

It is assumed that each node has at least one adjacent node that can act as a backup in case of failure. The ART algorithms are involved in choosing one of them as a backup node to re-route a packet through it when a failure occurs. The failure will occur ran- domly during simulation time in different positions along the primary path. In the net- work, the protocol starts to converge along the network and then ART algorithms will begin to operate when the routing protocol builds the routing table for each node on the topology.

## 4. ALGORITHM OVERVIEW

The ART algorithms were based upon the original routing table, which is computed by the Link State (LS) protocol to create a new backup path for all destinations. In this section, we discuss in detailed  the  basic concepts of the two new algorithms in wired networks by providing examples.

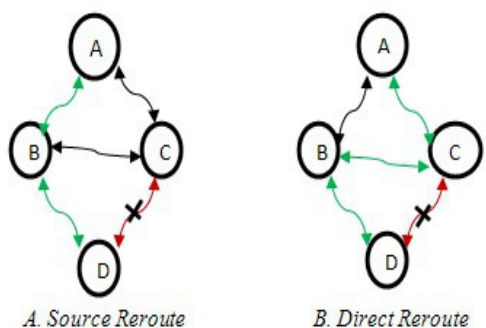

*A. Source Reroute*      *B. Direct Reroute*

|   | A | B | C | D |   |   | A | B | C | D |
|---|---|---|---|---|---|---|---|---|---|---|
| A | -- | B | C | C |   | A | -- | B | C | C |
| B | A | -- | C | D |   | B | A | -- | C | D |
| C | A | B | -- | D |   | C | A | B | -- | D |
| D | C | B | C | -- |   | D | C | B | C | -- |

Fig. 1: The primary and Backup Paths

### 4.1  ART REROUTE FROM SOURCE TO DESTINATION (ARTFP)

In the case of the re-route from source node, after the routing table has computed each node on the topology it sends small messages to all nodes that are adjacent to the disabled node connected to the primary path. The routing table makes nodes aware of the location of their neighbours. When the adjacent nodes receive these massages, they start to check hop-by-hop to the destination node. If there is no next hop connected to the primary path then they will inform the source that a backup path exists to the destination. When the adjacent node has checked the routing table to see if there is any node or link joined to any corresponding ones on the primary path, it will send an acknowledgement that informs the source node, saying, "I cannot be a backup node in case of failure". Hereafter, when the source node receives a negative acknowledgement it will neglect this node, and will check the other messages from other adjacent nodes to see if any node has a disjoint route to the destination. On the other hand, if the source node receives a "Yes" then it will add this node to the new backup routing table as a first next hop. However, if the source node receives a negative answer from all adjacent nodes that there is no disjoint path to the destination then the source node will send a message to the adjacent nodes (not along the primary path) to check whether their neighbours have a disjoint route to the destination. In this case, the adjacent node will receive an acknowledgement from its neighbours that inform it if there is a disjoint path

---

**Algorithm** 1 *AlternativePath* ART with Full Disjoint Path (ARTFP)

$G(V, E)$ is an oriented graph with two sets, a set of vertices $V$ and a set of edges $E$, where an edge $e = (v, u)$, $e \in E$, $v, u \in V$ is a connection from vertex $v$ to vertex $u$. A path $P$ is a set if edges $e_1, e_2, ..., e_n$, such that if $v, u, x \in V$, then $e_i = (v, x)$, $e_{i+1} = (x, u)$, for all

$1 \leq i \leq n - 1$.

1: procedure *AlternativePath*($T_r$)
2:   $T_r$: The routing table
3:   $V$: The vertex set in graph $G(V, E)$
4:   $\Gamma(v)$: The set of adjacent vertices to a vertex $v$
5:   $P_r(T_r, s, d)$: The path connecting the vertex $s$ to $d$ as in $T_r$
6:   $P_a(s, d)$: An alternative path such that $P_r(T_r, s, d) \cap P_a(s, d) = \varnothing$
7:   $SP$: The set of all generated alternative paths
8:   $q_{sub}$: A path
9:   $Q$: A queue of couple (path, vertex)
10:   $Enqueue$: Insert an element in a queue
11:   $Dequeue$: Removes an element from a queue
12:   $Front$: The element at the front of a queue
13:   $SP \leftarrow \varnothing$
14:   for all $s \in V$ do
15:       for all $d \in V$ do
16:         if $s = d$ then
17:           $q_{sub} \leftarrow \varnothing$
18:           $Q \leftarrow \varnothing$
19:           $Enqueue(Q, (q_{sub}, s))$
20:           while $Q = \varnothing$ and $p_a(s, d) = \varnothing$ do
21:             $(q_{sub}, x) \leftarrow Front(Q)$
22:             for all $k \in \Gamma(x)$ do
23:               $e \leftarrow (x, k)$
24:               if $(q_{sub} \cup e) \cap P_r(T_r, s, d) = \varnothing$ then
25:                 if $P_r(T_r, k, d) \cap P_r(T_r, s, d) = \varnothing$ then
26:                   $p_a(s, d) \leftarrow q_{sub} \cup e \cup P_r(T_r, k, d)$
27:                   $SP \leftarrow SP \cup p_a(s, d)$
28:                   break
29:                 else
30:                   $Enqueue(Q, (q_{sub} \cup e, k))$
31:                 end if
32:               end if
33:             end for
34:             $Dequeue(Q)$
35:           end while
36:           $Q \leftarrow \varnothing$
37:         end if
38:       end for

39: end for
40: return *SP*
41: end procedure

to the destination. If the acknowledgement is positive, the adjacent node will send a message to the source saying "Yes, I have a disjoint path via me and my neighbours". The source node will then add the adjacent node as a first and her neighbours as second hop in the new backup routing table [1]. Figure 1(A) shows how adjacent nodes can have a disjoint path to the destination without being connected to the primary path. With the primary path A–>C–>D, A sends a message to B to ask if there is any route to the destination D. B will check the next hop, which is C,D and because C is connected to the primary path it will be excluded by the adjacent nodes. Therefore, B concludes the next hop is D, which is the destination node and will then reply with an acknowledgement that I have a disjoint path. When link between C–>D fails, A will receive a notification about the failure and then will re-route the traffic directly through B to D.

In figure 2 we illustrate by example how ARTFP algorithm works to create a new backup routing table. We have a full topology with 11 nodes. The source node will send packets to the adjacent nodes {2 and 3} ignoring node 1, because node 1 is the first hop on the primary path. Node 2 will check if there is any disjoint path to the destination node but not via {5, 1, 6, 9} from the routing table. Node 2 will send an acknowledgement to inform the source whether it can act as a first backup hop .In figure 2, node 2 cannot be a backup hop because there is a common node in its path to the destination, which is node 6. Hereafter, the source node will check the other answers from other adjacent nodes. Node 3 will send acknowledgement to the source that, "I have a disjoint path to the destination without any nodes/links joined to the primary path". When the source node receives this answer, node 3 will add itself into the backup routing table as a first next hop in case of failure. On the other hand, if nodes {3 and 2} cannot be backup adjacent nodes and they do not have a disjoint path with the primary path to the destination, then the source node will send a packet for nodes {3 and 2} to check if they have an adjacent node that has a disjoint path to the destination. Hence, node 2 will send a packet to node 7 to ask if there is a disjoint path with the primary path to the destination. Hence, node 7 will check the routing table to discover if there is a path available. Detailed description ARTFP algorithm for rerouting from source to its destination is given in algorithm 1.

## 4.2 ART REROUTE TRAFFIC FROM NODE CONNECTED WITH FAILURE (ARTCP)

In this section, we explained the ARTCP algorithm to formulate how each node on the primary path re-routes traffic without waiting for the source node to receive a notification about the failure. In this algorithm, each node which is associated with a failure will re-route its traffic through its backup routing table, which is computed in advance. When a failure occurs, the node that is connected to the failed link will notice the malfunction before any node on the topology through loss of link signals by layer 2. This node will directly re-route its traffic to the destination. The dissimilarity between ARTCP and ARTFP, is that the former one constructs the backup path from the failed node on the primary path not from the source node which was the case for ARTFP. In ARTCP, each node on the primary path will seek for an adjacent node that can pass packets to the destination without returning the packets to any previous hop to avoid loop in the network. Each node on the primary path will send small packets to enquire from her adjacent if it has an available route to destination. The route should not have any node that has already

**Algorithm** 2 *AlternativePath* ART with Reroute from Node Connected with failure (ARTCP)

1: procedure *AlternativePath*($T_r$, $s$, $d$, *edges_to_avoid*)
2: $T_r$: The routing table
3: $V$: The vertex set in graph $G(V, E)$
4: $\Gamma(v)$: The set of adjacent vertices to a vertex $v$
5: $s$: The source vertex
6: $d$: The destination vertex
7: *edges_to_avoid*: A list of edges to exclude while implementing the alternative path
8: $pa(s, d)$: An alternative path such that *edges_to_avoid* $\cap Pa(s, d) = \emptyset$
9: $qsub$: A path
10: $Q$: A queue of couple (path, vertex)
11: *Enqueue*: Insert an element in a queue
12: *Dequeue*: Removes an element from a queue
13: *Front*: The element at the front of a queue
14: $pa(s, d) \leftarrow \emptyset$
15: if $s = d$ then
16: $qsub \leftarrow \emptyset$
17: $Q \leftarrow \emptyset$
18: $Enqueue(Q, (qsub, s))$
19: while $Q = \emptyset$ and $pa(s, d) = \emptyset$ do
20: $(qsub, x) \leftarrow Front(Q)$
21: for all $k \in \Gamma(x)$ do
22: $e \leftarrow (x, k)$
23: if $(qsub \cup e) \cap edges\_to\_avoid = \emptyset$ then
24: if $Pr(T_r, k, d) \cap edges\_to\_avoid = \emptyset$ then
25: $pa(s, d) \leftarrow qsub \cup e \cup Pr(T_r, k, d)$
26: break
27: else
28: $Enqueue(Q, (qsub \cup e, k))$
29: end if
30: end if
31: end for
32: $Dequeue(Q)$
33: end while
34: $Q \leftarrow \emptyset$
35: end if
36: return $pa(s, d)$
37: end procedure

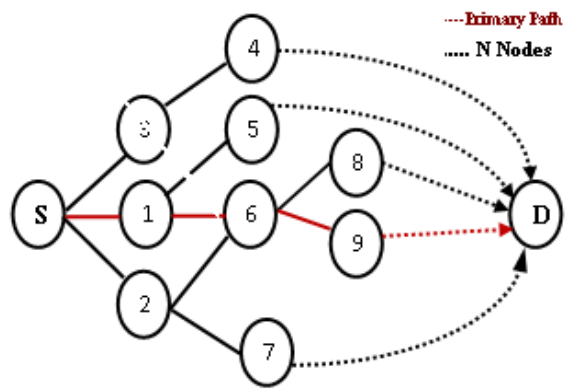

Fig. 2: The traffic Reroute

passed the traffic or is connected with any previous hop on the primary path. Link State Protocols will re-compute a new routing table for each node on the topology in order to make the source node re-route its traffic through new shortest path. During that time, the node that is connected to the failure will have a disjoint path to the destination with a guaranteed loop free in the network. According to figure 1 (B), the primary path that follows A->C->D, each node on the primary path will ask her neighbours if they have available route or not. In 1 (B), A, C sends a packets to B to check if it has a route to D. As we can see the failure occurred between C and D. Through the routing table B it can pass the traffic to D without returning packets to A. When failure occurs, C will reroute the traffic without the need to wait until A receive a notification and reroute the traffic after the routing protocol update the routing table. Detailed description of the ARTCP algorithm in rerouting from the failed node on the primary path to the destination is given in algorithm 2.

## 5 RESULTS

### 5.1 SIMULATION ENVIRONMENT

Network simulation (NS2) was performed to evaluate the performance of the proposed algorithms. We compared the simulation results of the link state protocol with ARTCP, ARTFP algorithms and without them. NS2 offers good support for wired topology in networks. We performed different network scenarios, each having a different number of nodes to demonstrate the effect of nodes or links failure during simulation time. Sim- ulations were ran 50 times with a series of random configured failures between source and destination. Failure can occur instantaneously during the simulation time in the net- work. In addition, we selected source and destination randomly. The duration time for each simulation was 50.0s. The CBR traffic was configured for all source nodes starts sending 200 kb/s from 1.0 to 50.0s. During that period, we caused numerous failures in different location for the links on the topology by creating variables [Random /Uniform]. This caused failure to occur arbitrarily. Hence, the links went down haphazardly. Each failure was recovered by its backup path, which was computed by the proposed algorithms .The simulation was configured for the link state protocol according to the table 1.

## 5.2 RESULT AND ANALYSIS

Simulating was repeated for all three algorithms ARTFP, ARTCP and the link state as the bench mark for identical network conditions. In ARTFP, the traffic will be re- routed along an alternative full disjoint path from a source to its destination when failure occurred. ARTCP will re-route the traffic from the node, which is connected directly to the failed link without the need to wait until the source node informs about it. When link state has updated the routing table then the traffic will pass again via a new primary shortest path.

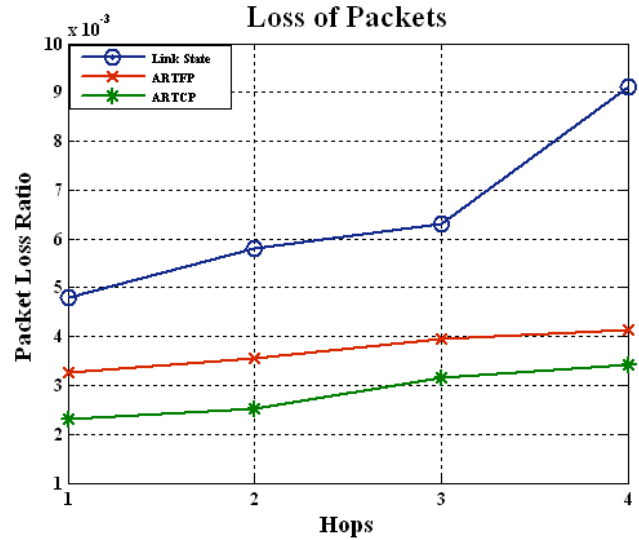

Fig. 3: Loss Of packets

Figure 3 shows the average lost packets versus the number of hops to the destination in three cases. The losses of packets were reduced in ARTFP and ARTCP algorithms because the traffic in ARTFP will re-route from the source when it receives a notification that a failure has occurred. Until the source node receives this notification the loss of packets will increase. On the other hand, the re-route from the node connected to the failure will reroute the traffic without need to wait until source node receives any notifications about it. Hence, this node will re-route the traffic via its backup routing table once it losses the connection with next hop node. During the process, the routing protocol will start to re-compute a new primary path to the destination with updating the routing table. When the LS protocol has completed all updates information for all nodes on the topology, then source node will pass the traffic via a new primary path. In the link state protocol, when failure occurs the source node will keep forwarding the traffic until it detects the failure and received a Link State Advertisement to con- firm it has occurred. Hence, the loop can be occurred between the nodes on the primary path. The node which is connected to the failure will return the packets to the previous hop according to the routing table. When the previous hop received the packets will return them again to that node according to the next hop in the primary routing table, because the earlier hop does not know about the failure yet. Therefore, the utilisation may exceed the size of link capacity, which increases loss of packets and degrades the network's performance.

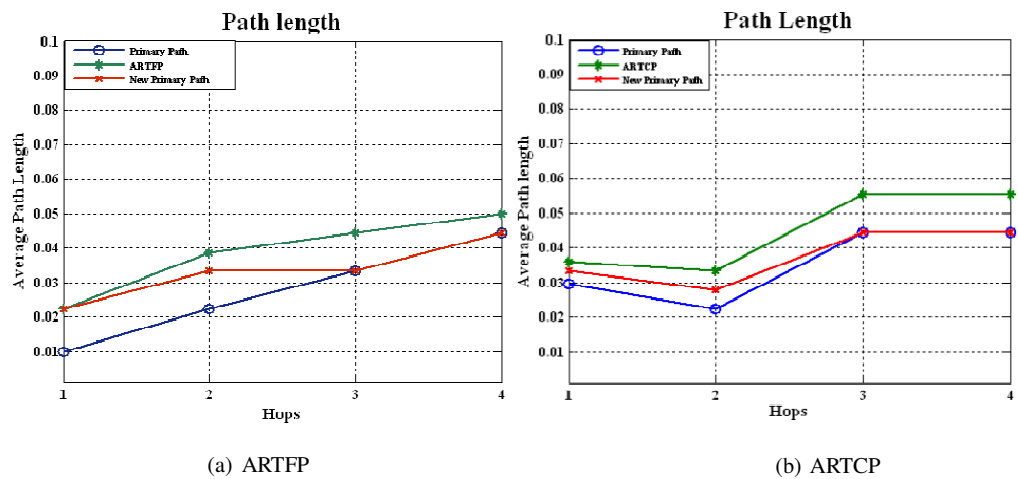

(a) ARTFP                  (b) ARTCP

Fig. 4: Average Path Length

Figure 4 (a) shows the average path length between source node and its destination in three cases; the primary path length which is computed by the link state protocol, backup path length which is computed by ARTFP algorithm and the new primary path length after the routing protocol has updated the new routing table. In ARTFP, the traffic will re-route from the disjoint path from source node to its destination, and we can see the ARTFP disjoint path is longer than the primary path and the new shortest path, which are computed by the routing protocol. ARTFP can choose path with more hops to guarantee loop free in the network. Figure 4 (b) shows the three cases for the ARTCP, Primary Path and New primary path after failure. Figure 4 (b), the average path length for ARTCP is longer than the other cases. ARTCP will be longer for all cases because the nodes on the primary path need to reroute the traffic via their adjacent nodes that require passing the traffic via extra hops to avoid loop by returning traffic to any node that the traffic has already passed from there.

Figure 5 shows the network throughput has improved where the link state protocol is configured with the two ART algorithms. Failure occurs in different locations on the primary path between source and destination; throughput is reduced based on the location of the failure from the source node. When a failure occurs far from the source, it will keep sending traffic until it receives a notification to inform it about the failure. In case re-routing from the node, which is connected directly to the failure (ARTCP), the loss of the link signals can lead the ARTCP to reroute the traffic faster than ARTFP and LS because both of them have to wait until they receive a notification about it. Hence, we can see that throughput has been improved compared to the ARTFP algorithm and LS protocol. In LS, when the source node received a notification then it has to wait until the LS re-computed a new primary path. In ARTFP, the traffic will reroute via the full disjoint backup path without the need to wait for the routing table to re-compute a new one.

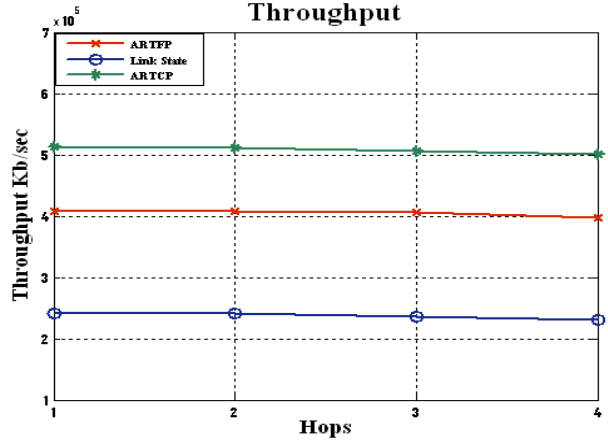

Fig. 5: Throughput.

Figure 6 illustrates the end-to-end delay between source and destination. The delay caused by the two ART algorithms of ARTFP and ARTCP is less than the link state protocol as we have shown. The ART algorithms are pre-active mechanisms and each node has a backup routing table that can re-route the traffic directly, therefore, when failure occurs the alternative path can be the same path for the link state in the new routing table. This will lead to reduced delay on the network. End-to-end delay will be reduced, because in the two cases, source re-route (ARTFP) and the node connected to the failure re-route ARTCP, they do not need to wait for the routing protocol to re-compute the new routing table and shortest path. They will re-route the traffic directly when they receive a notification about a failure via the backup routing table.

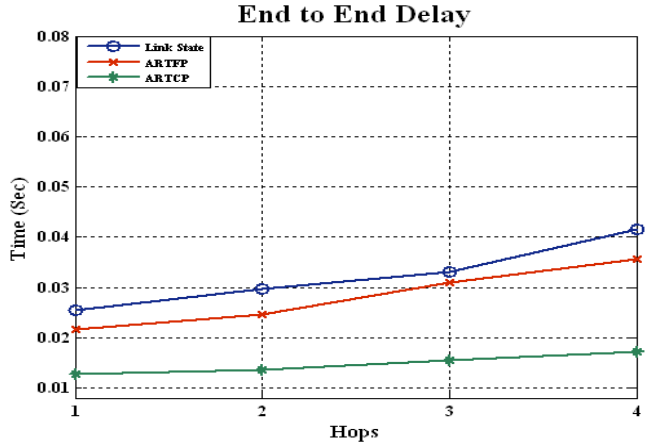

Fig. 6: End to End Delay.

Figure 7 shows the traffic load on the number of nodes in the network topology. This is with respect to the link state protocol with ART algorithms. The load on the link state with the ART algorithms is higher than the load for the link state alone. This is because ART algorithms lead all nodes on the topology flooding additional small packets to enquire from adjacent nodes about the availability of an alternative path to the destination. In addition, the ARTFP algorithm use a greater number of nodes than the link state protocol and ARTCP algorithm because these nodes will not join or have any nodes connected to the primary path. ARTCP, each node can find an alternative path without needing to send these small packets several times. In addition, ARTCP can choose the alternative path from its adjacent node, which does not need to send the packets to any previous node on the primary path to avoid loop in the network. How- ever, these small packets enquire from all adjacent nodes if there is an available route to the destination disjoint with the primary path. Therefore, the comparison shows that the ART algorithms do not generate high loads, which will degrade the network's performance. Additionally, the ART algorithms can cause a greater load on nodes compared to the link state protocol. This is because they used a disjoint backup path regardless of whether it is the shortest path.

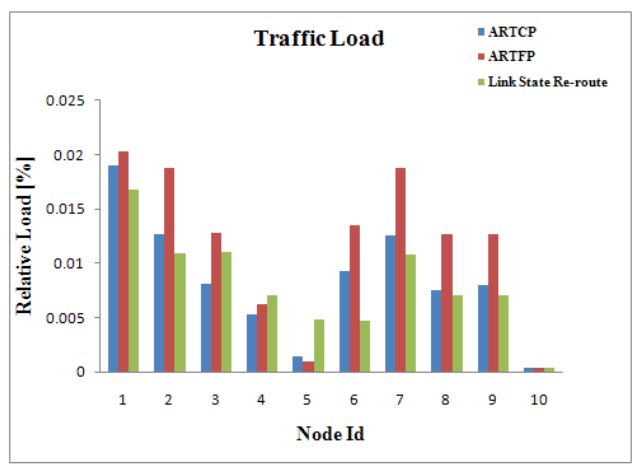

Fig. 7: Node Load

# CONCLUSION

Two new algorithms have been presented in this paper for computing an alternative path for each node on the network. The ART algorithms invest in the routing table, which is computed by a link state protocol to find an alternative disjoint path to the destination by creating a new backup routing table. Although ART algorithms can build a new backup routing table with shortest path, we have shown that backup paths that contain a few hops and in some cases they can choose the shortest path as in the new routing table. For real traffic, the result shows that our algorithm reduces the loss of packets and delay between source and destination node. In the future work, we will make the backup routing table contain both first and second shortest paths to the destination by choosing paths with minimum cost in order to create an optimal backup routing table.